\documentstyle[preprint,aps,epsfig]{revtex}

\newenvironment{natureabstract}{%
\begin{quote} \bf}
{\end{quote}}

\title{Bunching of Fractionally-Charged Quasiparticles Tunneling through High Potential Barriers}
\author{Comforti,~E., Chung,~Y.~C., Heiblum,~M., Umansky,~V. and Mahalu,~D.}
\address{Braun Center for Submicron Research, Department of Condensed
 Matter Physics,\\ Weizmann Institute of Science, Rehovot 76100, Israel}
\date{\today}

\begin{document}

\maketitle

\begin{natureabstract}
Shot noise measurements were recently exploited to measure
the charge of the quasiparticles in the Fractional Quantum
Hall (FQH) regime. For fractional filling factors $\nu=1/3$
and $2/5$ of the first Landau level, fractional charges
$q=e/3$ and $e/5$, respectively, were
measured\cite{rafi,glattli,misha}.  We investigate here the
interaction of $e/3$ quasiparticles with a strong
backscatterer and find unexpected results. When a weak
backscatterer is introduced in the path of an otherwise
noiseless current of quasiparticles, stochastic
\textit{partitioning} of the quasiparticles takes place and
shot noise proportional to their charge appears.
Specifically, at $\nu=1/3$, noise corresponding to $q=e/3$
appears.  However, the measured charge increases
monotonically as backscattering becomes stronger,
approaching asymptotically $q=e$~\cite{kf,tim}.  In other
words, only \textit{electrons}, or alternatively, three
\textit{bunched} quasiparticles, tunnel through high
potential barriers when impinged by a noiseless current of
quasiparticles.  Here we show that such bunching of
quasiparticles by a strong backscatterer depends on the
average occupation (\textit{dilution}) of the
\textit{impinging} quasiparticle current.  For a very dilute
impinging current, bunching ceases altogether and the
transferred charge approaches $q=e/3$.  These surprising
results prove that a sparse beam of quasiparticles, each
with charge $e/3$, tunnel through high potential barriers,
originally thought to be opaque for them.
\end{natureabstract}

Already in 1918 W. Schottky determined the charge of the
electron by measuring the average square of the current
fluctuations, $S=\langle i^2 \rangle$ , resulting from the
stochastic emission of electrons in a vacuum tube - naming
it \textit{shot noise}\cite{schottky} .  His famous
expression $S=2qI$ is a result of independent random events
obeying Poisson distribution. Here, $S$ is the spectral
density of the fluctuations (in units of A$^2$/Hz), $q$ the
charge of the particle and $I$ the average current. Similar
experiments were implemented~\cite{rafi,glattli,misha} in the
FQH regime\cite{fqhe}, verifying the existence of
fractionally charged quasiparticles\cite{laughlin}.  A
partially transmitting constriction, \textit{Quantum Point
Contact} (QPC), served in these experiments as an adjustable
potential barrier in the path of the current, thus
partitioning the transmitted charges. This random process is
described by a binomial distribution, resulting with
$S=2qI(1-t)$ at zero temperature, with $t\in [0,1]$ being the
transmission coefficient of the QPC\cite{kf,khlus,lesovik}.
In the weak backscattering regime, the quasiparticles were
found to traverse the QPC independent of each other and the
measured charge was $q=e/3$ at filling factor $\nu=1/3$ and
$q=e/5$ for $\nu=2/5$~\cite{rafi,glattli,misha}.  As
backscattering gets stronger, the tunneling of individual
quasiparticles becomes correlated, and in the limit of a
\textit{pinched} QPC and $\nu=1/3$ three quasiparticles were
found to group together in order to tunnel through the
barrier. Obviously, Schottky's formula is inapplicable for
\textit{correlated} (or \textit{bunched})
\textit{quasiparticles}, but it can still be used to
characterize the system with the effect of bunching being
incorporated into an effective charge $q(t)$. Hence, the
noise for a pinched QPC becomes \textit{electronic
like}\cite{kf}, namely, with an effective charge $q=e$.
Moreover, a nearly universal behavior was found for the
evolution of the effective charge $q(t)$~\cite{tim}, starting
at $q($open QPC$)=e/3$ and monotonically increases toward
$q($pinched QPC$)=e$.

Here we explore the \textit{bunching properties} of a
pinched QPC when a \textit{sparse} beam of $e/3$
quasiparticles impinges upon it. In other words, when there
are not enough quasiparticles in close proximity to bunch
into an 'electron', we may ask the following questions: (a)
Will the barrier become opaque? (b) Will quasiparticles be
'borrowed' from lower-energy states to fill in for the
missing ones in the beam?  Or, (c) perhaps the bunching
condition will be relaxed and individual quasiparticles will
transverse the barrier? Theory does not provide yet answers
to these questions.

Samples were fabricated in a high-mobility two-dimensional
electron gas embedded in a GaAs-AlGaAs heterostructure.
Measurements were done in a strong magnetic field $B\approx
13$~T and a fractional filling factor $\nu=1/3$ in the FQH
regime. Vanishing of the longitudinal resistivity $\rho_{xx}$
assures that the (net) current is flowing chiraly along the
edges of the sample in \textit{edge states}.  This allows
measurements in \textit{multi-terminal geometries} shown in
Fig. 1.  Two techniques are employed in order to partition
the quasiparticle beam, hence making it \textit{sparse} or
\textit{dilute}, before it impinges on the pinched QPC2.  A
straightforward scheme, shown in Fig. 1a, utilizes a
noiseless current $I_{inc}$ that impinges on a relatively
\textit{open} QPC1 and partially scatters toward QPC2
(although in this case the small \textit{reflection
coefficient} of QPC1 is responsible for the dilution of the
current, for uniformity we stick to the notation of
\textit{transmission coefficient} $t_1\rightarrow 0$). Most
of the current continues toward
\textit{\textbf{D}}$\mathbf{_1}$ while the scattered part is
a very dilute beam of quasiparticles with charge
$q_1=e/3$~\cite{rafi,glattli,tim} and dilution determined by
$t_1$. Much of that dilute current is reflected back by QPC2
toward drain \textit{\textbf{D}}$\mathbf{_2}$ and a small
part $t_2$ is transmitted.  A fraction $t_1t_2$ of the
incident current reaches the amplifier.  This method cannot
be applied, however, to achieve a moderately sparse beam of
quasiparticles since a partly pinched QPC1 would lead to
bunching of quasiparticles and to an \textit{effective
charge} $q_1>e/3$~\cite{tim}.  Hence, we employ also a
geometry shown in Fig. 1b.  Here, the incident current is
being partitioned by transmission through a cascade of weakly
backscattering QPCs (for each $t_i\rightarrow 1$), feeding
QPC2 with current of quasiparticles with arbitrary dilution
$t_1=\prod_i t_i$.  The fact that the partitioned charge by
this method is $e/3$ is not obvious and a detailed
study\cite{prl} was needed to prove that a beam of
quasiparticles is indeed produced. We also tested, via
detailed noise measurements\cite{prl}, whether dilute
quasiparticles suffer \textit{intra-edge scattering} and
subsequent equilibration during transport along the device
edges. Equilibration establishes a new chemical potential and
increases the occupation of each state below the chemical
potential - hence, modifying the dilution of the beam.  As
shown in Ref. 11 such equilibration does not take place in
our devices.

Using one of the two methods depicted in Fig. 1 we create a
\textit{noisy} beam of quasiparticles with charge $e/3$,
which is being further partitioned by QPC2.  The noise at
\textit{\textbf{A}} is measured with a spectrum analyzer
after amplification by a cooled amplifier.  The amplifier,
being placed near the sample, has a very low current noise
at its input, $\langle i^2_{amp}\rangle=1.5\times
10^{-28}$~A$^2$/Hz, when operating with bandwidth of 30 kHz
and center frequency $f_0\approx 1.5$~MHz. The value of
$f_0$, chosen well above the cutoff of the ubiquitous $1/f$
noise, is determined by a resonance of $LC$ circuit, with $C$
determined by the capacitance of the coaxial cable
connecting the sample and the amplifier and $L$ by an added
superconducting coil\cite{rafi}. Reflected currents flow into
the grounded terminals \textit{\textbf{D}} and
\textit{\textbf{T}}, leading to a constant \textit{input}
(at \textit{\textbf{S}}) and \textit{output} (at
\textit{\textbf{A}}) conductance $G=e^2/3h$ - independent of
the transmission of the QPCs.  This makes both the sample's
equilibrium noise ($4k_BTG=5\times 10^{-29}$~A$^2$/Hz) and
the sample-dependent amplifier's noise ($\langle
i^2_{amp}\rangle/G^2$) independent of QPCs' transmission,
allowing their subtraction from the measured noise (for
comparison, the magnitude of the shot noise at
\textit{\textbf{A}} is typically in the $10^{-30}$~A$^2$/Hz
range).

The configuration in Fig. 1a can be analyzed by means of
\textit{superposition}\cite{yamamoto}.  Consider first the
'injector' QPC1, characterized by transmission
$t_1\rightarrow 0$ toward QPC2 and partitioned charge $e/3$.
Being a stochastic element, it generates (at zero
temperature) noise $2(e/3)I_{inc}t_1(1-t_1)$, with
$I_{inc}t_1$ the transmitted current, impinging on QPC2. This
noise is attenuated with a factor $t^2_2$ by QPC2, resulting
with a contribution of QPC1 to the total noise:

\begin{equation}\label{s1}
S_1=2(e/3)I_{inc}t_1(1-t_1)\cdot t^2_2 \ .
\end{equation}

\noindent Consider now QPC2, characterized by transmission
$t_2$ and charge $q_2$ when impinged by a \textit{noiseless}
current $I_{imp}$ of $e/3$ quasiparticles.  It produces
noise $2q_2I_{imp}t_2(1-\widetilde{t_2})$, where
$I_{imp}t_2=I_2$ is the transmitted current and
$\widetilde{t_2}=t_2\frac{e/3}{q_2}$ denotes the effective
transmission for charge $q_2$ quasiparticles. This
transmission $\widetilde{t_2}$ is determined self
consistently with the charge $q_2$ in order to maintain the
measured conductance of QPC2\cite{tim}. We stress that even
though the current $I_{imp}=I_{inc}t_1$ is noisy we still use
the above expression to calculate the noise generated by
QPC2, since the noise in $I_{imp}$ was already taken into
account in Eq.~(1).  The added contribution of QPC2 is
therefore:

\begin{equation}\label{s2}
S_2=2q_2I_{inc}t_1t_2(1-\widetilde{t_2}) \ .
\end{equation}

\noindent The total noise in \textit{\textbf{A}} is then
$S_1+S_2$.  The correctness of this analysis can be validated
in the limit of a constant charge (say, $e/3$):
$S_1+S_2=2(e/3)I_{inc}t_1t_2(1-t_1t_2)$, with
$t_{tot}=t_1t_2$ being the total transmission from
\textit{\textbf{S}} to \textit{\textbf{A}} - the standard
expression for binomially distributed process. In the
experiment we use the expression for $S_1+S_2$ in order to
determine the charge $q_2$ partitioned by QPC2.  In the limit
where both $t_1$ and $t_2$ are small $S_1$ and $S_2$ are of
order of $O(t_1t^2_2)$ and $O(t_1t_2)$ , respectively, with
the first much smaller than the second.  Hence, the measured
noise is dominated by the contribution of the pinched QPC2:

\begin{equation}\label{s_limit}
S\approx 2q_2I_{inc}t_1t_2=2q_2I_2 \ .
\end{equation}

We verify first that the noise produced by a pinched QPC,
when fed with a quiet current, corresponds as expected to
$q\approx e$.  We find results similar to these in Ref. 5
with an example given in Fig. 3b.  For $t_1=1$, hence feeding
QPC2 with a noiseless current, and $t_2\approx 0.1$, we
measure indeed charge $e$.  We then partition the incident
current by setting $t_1<1$, hence impinging a noisy current
of quasiparticles on QPC2 with $t_2\approx 0.1$.  The noise
seen in Fig. 2a corresponds to an average state-occupation
$t_1\approx 0.7$ and than in Fig. 2b to $t_1\approx 0.2$.
Calculating the expected noise\cite{tim} we take into account
the finite temperature of the electrons ($T\approx 65$~mK)
and the energy (or current) dependence of the total
transmission $t_{tot}=t_1t_2$ (current dependent
transmissions are shown in the insets of Fig. 2). The
average current is being varied over a large enough range,
with the voltage $V$ satisfying $q_2V\gg k_BT$, to allow the
noise to reach the linear regime. Nice agreement is found
between the data and the independent particle model for
$q_2=0.9e$ and $t_1=0.7$ (lightly diluted current) and for
$q_2=0.55e$ and $t_1=0.2$ (highly diluted current).

A more striking example of the effect of beam dilution is
demonstrated in Fig. 3, where the range of current $I_2$ is
kept constant for different values of dilution.  Obviously,
a higher source voltage is required to obtain the same
current $I_2$ when the current is more dilute.  In comparison
with the measured \textit{electron charge} for a noiseless
impinging current (Fig. 3b), a highly dilute current
($t_1\approx 0.1$) impinging on the pinched QPC2 is found to
produce a small charge $q_2=0.45e$ (Fig. 3a) - slightly
above the quasiparticles charge.

Figure 4 summarizes the dependence of $q_2$ on the dilution
$t_1$ of the impinging current on QPC2.  Two examples,
$t_2=0.1$ and $t_2=0.25$, are chosen, with corresponding
charge $e$ and $0.75e$, respectively, for a noiseless
impinging current. The more dilute the impinging current is
($t_1\rightarrow 0$), the smaller is the effective charge
$q_2$ - approaching asymptotically $~e/3$. The unavoidable
conclusion is that \textit{bunching of quasiparticles} is
not an essential mechanism for quasiparticles transfer
through high potential barriers! Bunching takes place
\textbf{only} when the incoming states are highly occupied.

Before we conclude we may also ask how is the
\textit{transmission} of QPC2 affected by the dilution of
the impinging quasiparticles.  Present theory assumes only a
noiseless current approaching a constriction within the
framework of the Luttinger model.  Also here we find counter
intuitive results.  In the linear regime, where the source
voltage is small enough to keep the transmission almost
energy independent (Fig. 5a) the transmission $t_2$ is
independent of dilution (although the source voltage for the
dilute current is some ten times larger).  This can be
compared with the case where the same source voltage range
is kept (Fig. 5b). Here, the transmission $t_2$ of the
noiseless current strongly depends on voltage (approaching
unity at $V>50$~$\mu$V).  Since equilibration of
quasiparticles had been ruled out\cite{prl}, we conclude
that the non-linearity of the pinched QPC depends strongly
on the quasiparticle \textit{current} and less on the
quasiparticle \textit{energy}. This rules out that the
potential profile of the barrier in the QPC is responsible
for non-linearity of the current. Moreover, the
insensitivity of the transmission $t_2$ in the linear regime
to the dilution of the impinging quasiparticle beam suggests
equal probabilities of tunneling for a single quasiparticle
and for bunched quasiparticles. In other words, noise and
transmission measurements show that quasiparticles can
transfer, with the same \textit{ease}, either \textit{one by
one} or \textit{bunched in groups}.  Their bunching depends
on the transparency of the barrier and on the preparation of
the quasiparticle beam.  It is now for theory to explain
such a bizarre effect.
\\
\\

\noindent \textbf{Acknowledgements:}

\noindent The work was partly supported by the Israeli
Academy of Science and by the German-Israel Foundation (GIF)
grant.  We thank A. Yacoby, A. Stern and Y. Levinson for
helpful discussions.
\\
\\
\textbf{Correspondence and requests for materials should be
addressed to M.H. (e-mail: heiblum@wisemail.weizmann.ac.il).
}\\
\\
\textbf{Captions}
\\
\textbf{Fig. 1.}  Schematic and actual representations of
quasiparticles \textit{injector} followed by a quasiparticle
\textit{filter}, both made of quantum point contacts, QPC1
and QPC2, respectively.  \textbf{(a)}~An injector made of a
relatively open QPC1 partitions an incident noiseless (dc)
current, injected from terminal \textit{\textbf{S}}.  The
scattered part ($t_1$), composed of a dilute beam of
quasiparticles, impinges on a pinched QPC2.  The resulting
noise measured by a cooled, low-noise, amplifier at terminal
\textit{\textbf{A}} (see Ref. 1). The intermediate drain
\textit{\textbf{D}}$\mathbf{_2}$ prohibits
multiple-reflections, and the grounded terminal
\textit{\textbf{T}} is used to fix the \textit{output}
impedance of the sample and make it independent of QPC
settings.  \textbf{(b)}~An alternative scheme, suitable for
producing a moderately dilute current, invokes a cascade of
weakly backscattering QPCs \textit{transmitting} a dilute
quasiparticle beam (see Ref. 11).  \textbf{(c)}~A photograph
of the actual device in the vicinity of the QPCs, formed by
metallic gates (light gray regions) deposited on the surface
of the GaAs-AlGaAs heterostructure embedding a two
dimensional electron gas some 0.1~$\mu$m below the surface.
Electron mobility is $2\times 10^6$ cm$^2$/Vs and areal
carrier density is $1.1\times 10^{11}$ cm$^{-2}$, both
measured at 4.2~K in the dark.  The solid arrows correspond
to the direction of current in configuration (a), while the
other QPCs on the right (with current flow denoted by dotted
arrows) are used when configuration (b) is employed.  Ohmic
contacts (serving as \textit{\textbf{S}},
\textit{\textbf{D}}, \textit{\textbf{T}},
\textit{\textbf{A}}) are outside the frame of the picture.
\\
\\
\textbf{Fig. 2.}  Noise and transmission measurements of the
pinched QPC2 (with transmission $t_2\approx 0.1$ at zero
bias) for two different values of dilution of the impinging
current: $t_1=0.7$~\textbf{(a)} and $0.2$~\textbf{(b)}.  In
the main graphs the measured noise is plotted against the
transmitted current, together with the theoretical
prediction of the independent particles model at a finite
temperature.  The intermediate curve in each graph
represents the best fit to an arbitrary charge $q_2$.
Various transmission coefficients, measured simultaneously
with the noise, are shown in the insets against the incident
(noiseless) current.  Each inset shows the dilution level
$t_1$ generated by the QPC1 \textit{injector}, the
transmission $t_2$ of the pinched QPC2 \textit{in response
to the dilute impinging current}, and the total transmission
$t_{tot}$. Notice that the sensitivity of $t_2$ to the
current depends on the dilution level of the impinging
current.
\\
\\
\textbf{Fig. 3.}  Comparison of the charge characterizing
the pinched QPC2 for two extreme cases of the impinging
current: not diluted (noiseless) and highly dilute,
\textit{keeping the same transmitted current}.
\textbf{(a)}~The noise produced by the pinched QPC2 when fed
by a highly dilute impinging current, $t_1\approx 0.1$,
corresponds to a quasiparticle charge $q_2\approx 0.45e$.
The inset shows the current-dependent transmission $t_1$
(level of dilution) and the transmission $t_2$ (is fairly
current independent for such a dilute beam).
\textbf{(b)}~The noise produced by the pinched QPC2 when fed
by a noiseless current corresponds to almost an
\textit{electronic} charge.  The inset verifies the charge
$q_2=e$ by measuring the noise over considerably wider range
of transmitted current. Clearly, the charge characterizing
QPC2 depends not only on the potential barrier height but
also on the average occupation of the states (dilution) of
the impinging current.
\\
\\
\textbf{Fig. 4.}  Evolution of the effective charge $q_2$
that characterizes the pinched QPC2 in response to different
values of dilution $t_1$ of the impinging current (extracted
from curves similar to that in Fig. 3a).  Results of three
different measurements are shown - two complementary sets of
data for $t_2=0.1$, with dilution produced by backscattering
of a single QPC1 (Fig. 1a) and by transmission through five,
relatively open, QPCs (Fig. 1b), and one set with $t_2=0.25$.
In the case of $t_2=0.25$ only the extreme points are shown
in order to simplify the graph.  The dashed lines are only
guide to the eye.  Evidently, as the current impinging on
the pinched QPC2 becomes more dilute, the charge drops from
its original value toward $~e/3$ in the limit of very high
dilution.  We conclude that individual, very sparse,
quasiparticles tunnel through a pinched QPC - originally
thought to be opaque for them.
\\
\\
\textbf{Fig. 5.}  Dependence of the transmission $t_2$ of
the pinched QPC2 on the dilution $t_1$ of the impinging
current. \textbf{(a)}~Measurements in the linear regime.  The
transmission $t_2$ of a highly dilute impinging current
($t_1\approx 0.1$, solid curve) and that of a noiseless
current ($t_1=1$, dashed curve), with the same
\textit{transmitted current} range kept in both cases.  The
applied source voltage, on the other hand, reaches a maximal
value of $~170$~$\mu$V for the noisy impinging current but
only $~16$~$\mu$V in the noiseless case. Nevertheless, the
transmissions in both cases are similar.
\textbf{(b)}~Measurements in the non-linear regime.  Similar
measurements to (a) but the same applied \textit{source
voltage} range is kept in both cases (the noiseless current
is some ten times larger for the same voltage).  The
transmission is found to be strongly current dependent when
the impinging current is noiseless.
\pagebreak
\begin{natureabstract}
\epsfxsize=13 cm \epsfbox{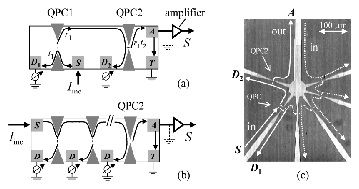}
\vspace{8 cm}
\begin{center}
Comforti \textit{et al.} - Figure 1
\end{center}
\pagebreak

\epsfxsize=13 cm \epsfbox{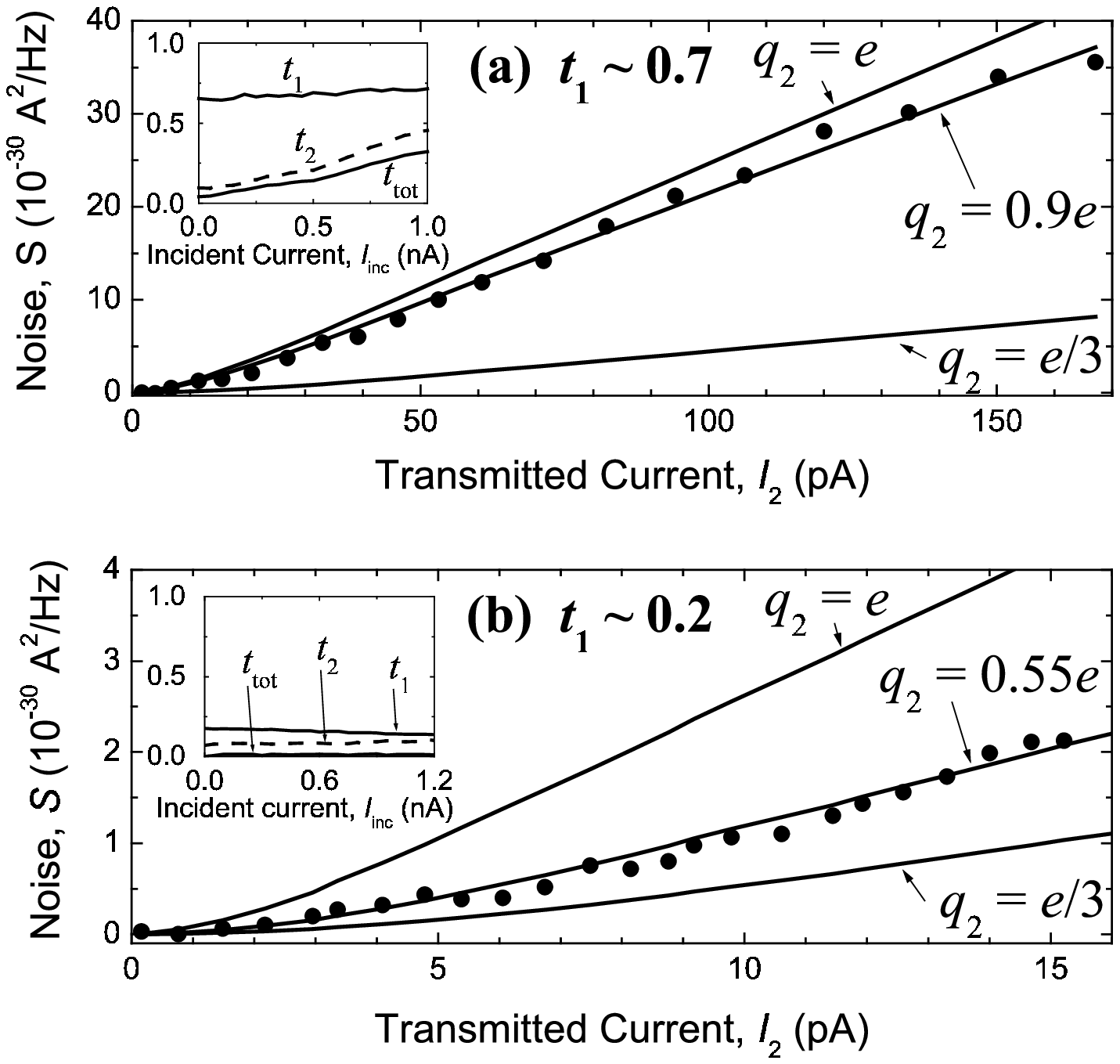}
\begin{center}
\vspace{3 cm} Comforti \textit{et al.} - Figure 2
\end{center}
\pagebreak

\epsfxsize=13 cm \epsfbox{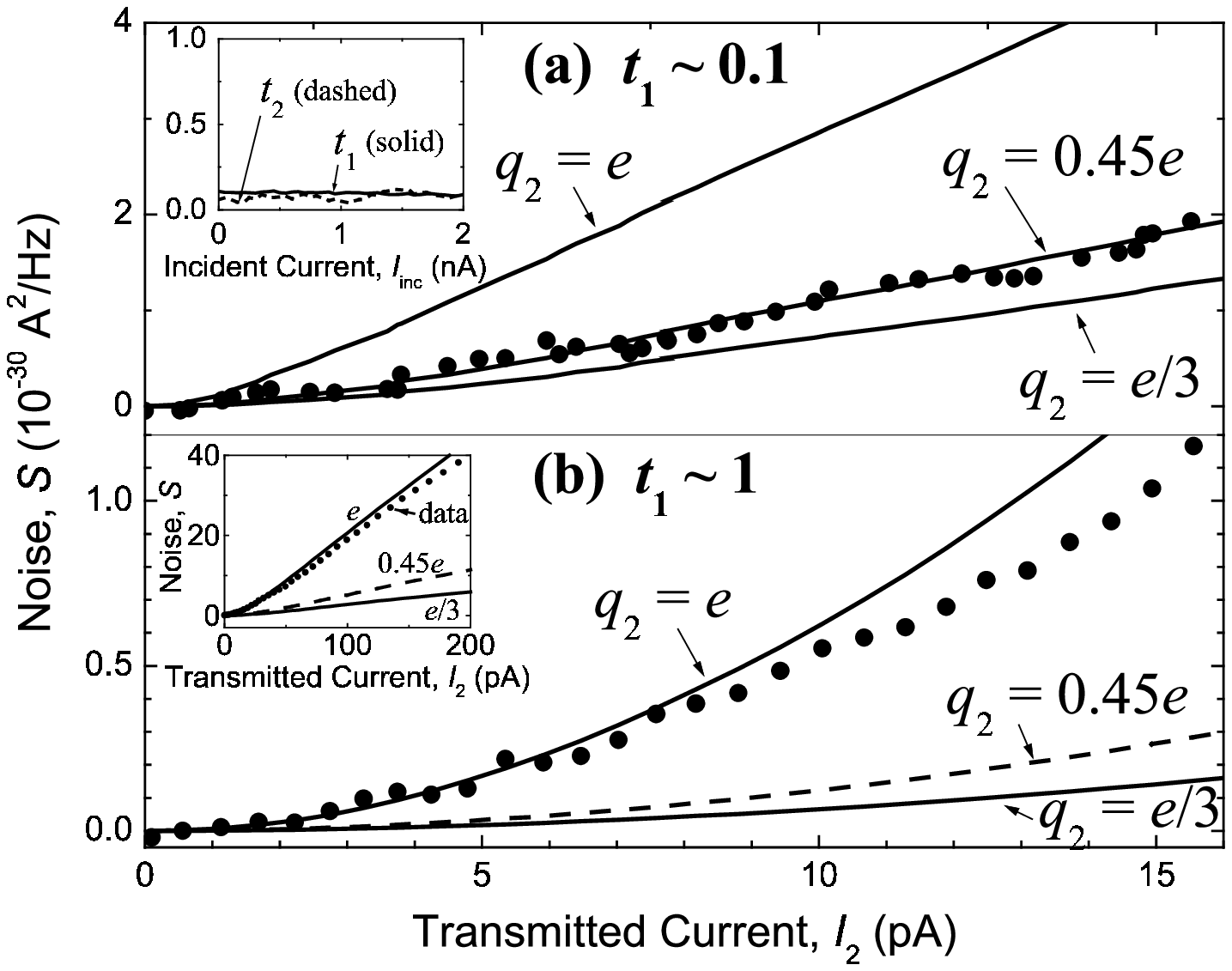}
\begin{center}
\vspace{5 cm} Comforti \textit{et al.} - Figure 3
\end{center}
\pagebreak

\epsfxsize=13 cm \epsfbox{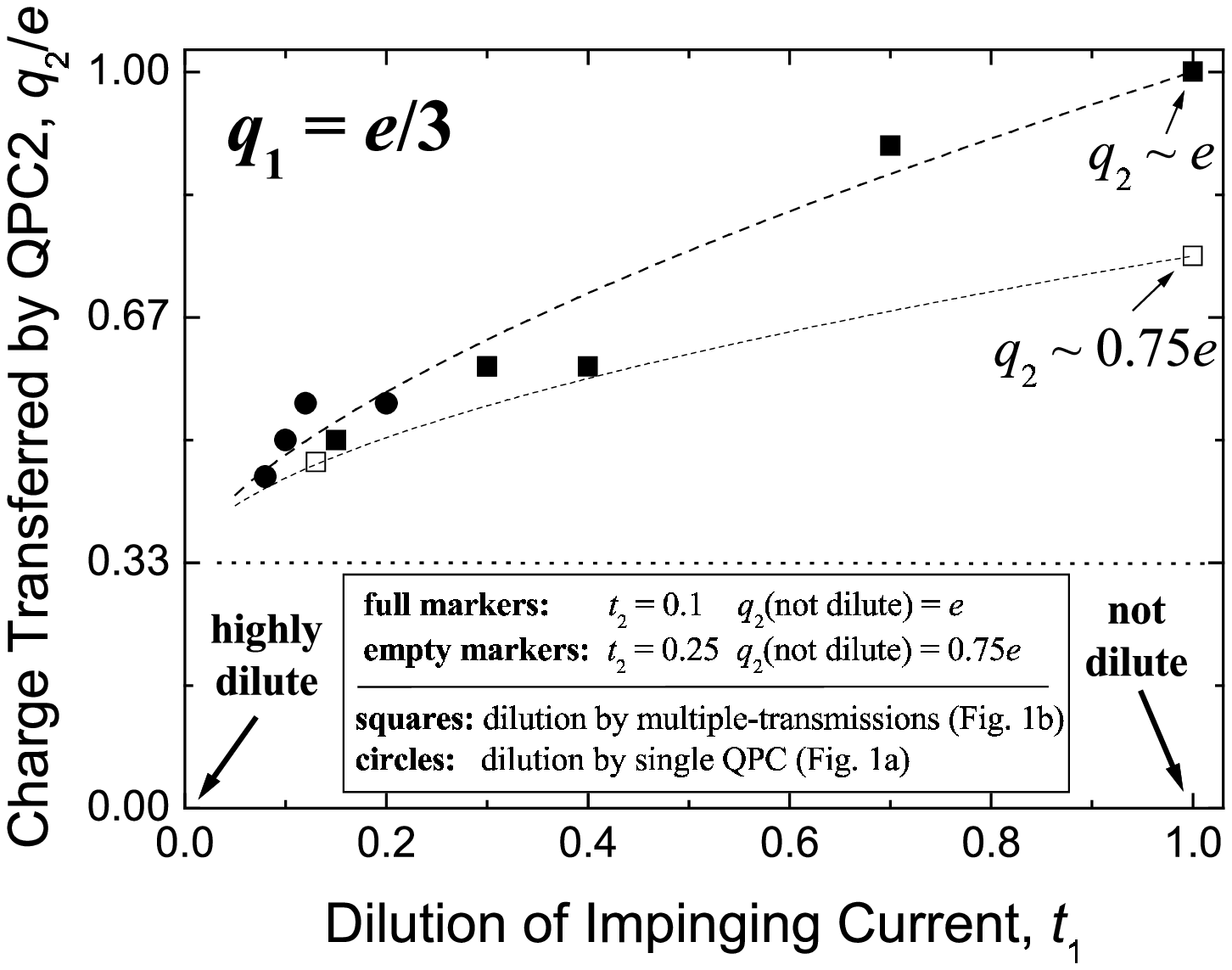}
\begin{center}
\vspace{5 cm} Comforti \textit{et al.} - Figure 4
\end{center}
\pagebreak

\epsfxsize=13 cm \epsfbox{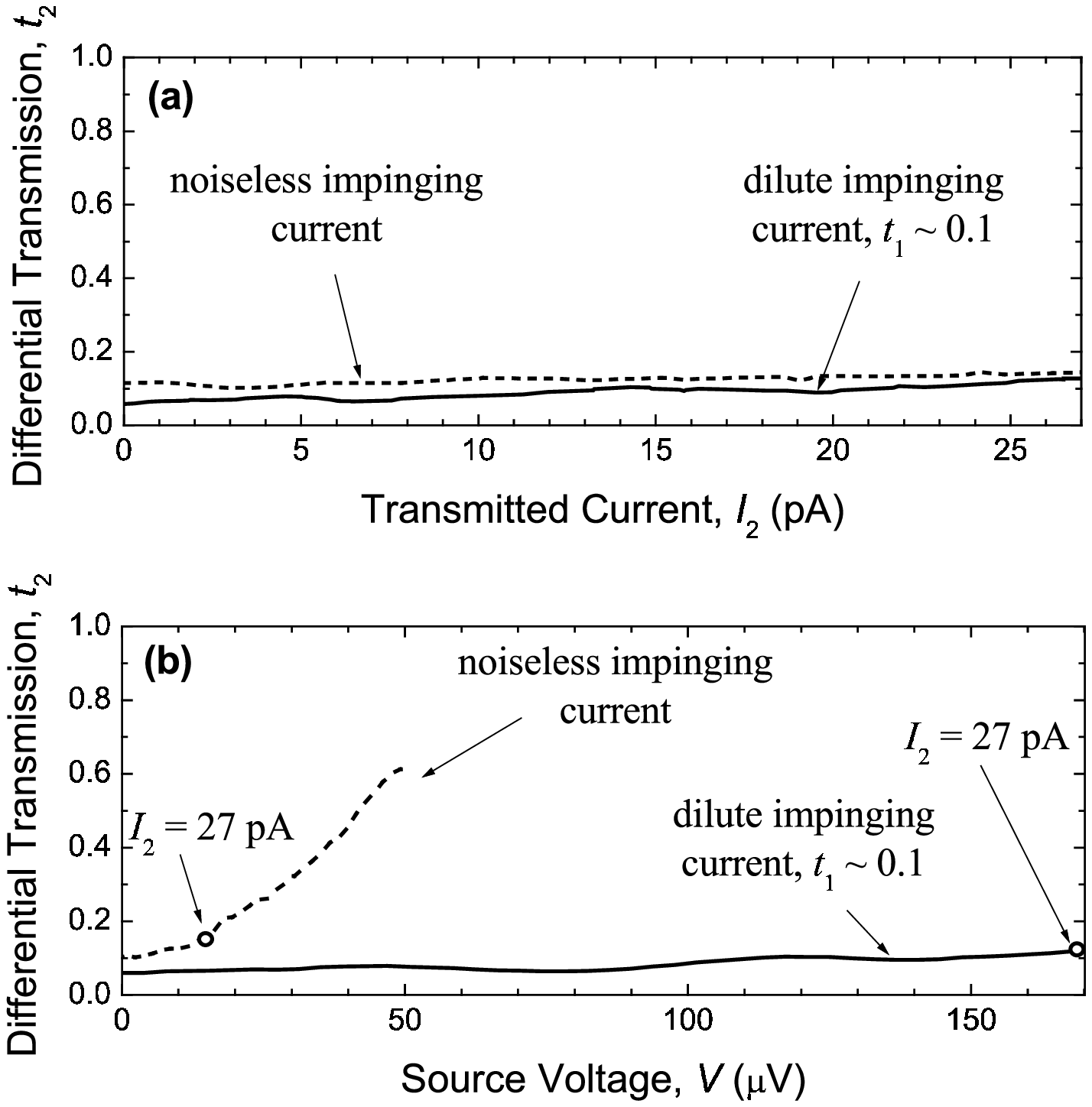}
\begin{center}
\vspace{2 cm} Comforti \textit{et al.} - Figure 5
\end{center}
\pagebreak

\end{natureabstract}
\end{document}